# Chalcogenide hyperbolic metamaterial with switchable negative refraction

Harish N. S. Krishnamoorthy,[1][*] Behrad Gholipour,[2][*] Nikolay I. Zheludev,[1,2] and Cesare Soci[1][†]

[1]Center for Disruptive Photonic Technologies, TPI, SPMS, Nanyang Technological University, Singapore 637371

[2] Optoelectronics Research Centre & Centre for Photonic Metamaterials, University of Southampton, SO17 1BJ, Southampton, UK

[†]Corresponding author: csoci@ntu.edu.sg

**Chalcogenide glasses are exceptional materials that show a wide contrast in optical properties upon phase change and are highly researched for reconfigurable electronic and nanophotonic devices. Here, we report the first proof-of-concept demonstration of a non-volatile, switchable hyperbolic metamaterial based on a chalcogenide glass. By using the $Ge_2Sb_2Te_5$ (GST) alloy as one of the components of a multilayered nanocomposite structure and exploiting its phase change property, we demonstrate a hyperbolic metamaterial in which the Type-I hyperbolic dispersion ($\varepsilon_\perp < 0$, $\varepsilon_\parallel > 0$) can be switched from the near infrared to the visible region. This opens up new opportunities for reconfigurable device applications, such as imaging, optical data storage and sensing.**

[*]These authors contributed equally to this work.

Metal-dielectric nanocomposite structures have attracted considerable interest over the last decade owing to their ability to support electron charge oscillations, known as surface plasmon polaritons. An example of such systems are hyperbolic metamaterials (HMMs), strongly anisotropic systems where the coupling between multiple surface plasmon polariton modes gives rise to a characteristic hyperbolic optical isofrequency surface, which supports electromagnetic states with large wave vectors [1]. These peculiar properties lead to unusual effects such as negative refraction [2–4] as well as nominally infinite photonic density of states [5], and have been exploited for a variety of applications such as optical imaging beyond the diffraction limit [6,7], enhancing spontaneous emission rates of quantum emitters [8–10], efficient heat transfer [11,12], nanoscale indefinite cavities [13,14], and optical sensing [15]. Unlike resonant systems such as microcavities and nanoscale resonator arrays, hyperbolic metamaterials exhibit their extraordinary electromagnetic properties over a wide spectral range, thereby making them useful for applications that require broadband response.

The principal characteristics of an HMM such as the nature of the hyperbolic dispersion and its onset frequency are determined by the constituent metal and dielectric materials as well as their relative fill-fraction, and usually cannot be changed once the HMM is fabricated. The ability to dynamically tune the topology of the optical isofrequency surface is significant as it can lead to the realization of tunable hyperlenses, imaging systems, and sensors. Moreover, future optical networks will require a new generation of adaptable integrated nanophotonic devices with functions such as optical switching and mode (de)multiplexing. Consequently, phase change materials are gathering growing interest to realise reconfigurable photonic devices. In this pursuit, phase change materials such as vanadium dioxide and chalcogenide glasses have emerged as unique platforms for realising optically reconfigurable metamaterials [16], metasurfaces [17–21], and volatile tunable HMMs [22].

In particular, chalcogenide compounds (binary and ternary sulphides, selenides and tellurides) are an exceptionally adaptable phase change material base, thanks to their compositionally-controlled high-index and low optical losses throughout a broad spectral range, from the visible to the far-infrared. These, combined with their non-volatile, reversible switching properties, which are easily triggered by thermal, electrical or optical means, have made them the materials of choice for a range of photonic devices with applications in mid-infrared sensing, integrated optics and ultrahigh-bandwidth signal processing [23–25]. Notably, phase-change properties of chalcogenides – reversible transitions between amorphous and crystalline states with starkly different optical (refractive index) and electronic (conductivity) characteristics, have been exploited for decades in optical data storage and more recently in electronic phase-change random access memories [26,27].

Here we utilise the chalcogenide alloy $Ge_2Sb_2Te_5$ (GST) as the non-volatile switchable constituent of a multilayer HMM. By exploiting the reversible amorphous to crystalline phase change transition of GST, we demonstrate the first chalcogenide-based reconfigurable HMM. As a proof-of-concept, we show that the negative refraction of an incident TM polarized light beam can be switched from the near infrared (IR) spectral region to the visible, as shown schematically in Figure 1.

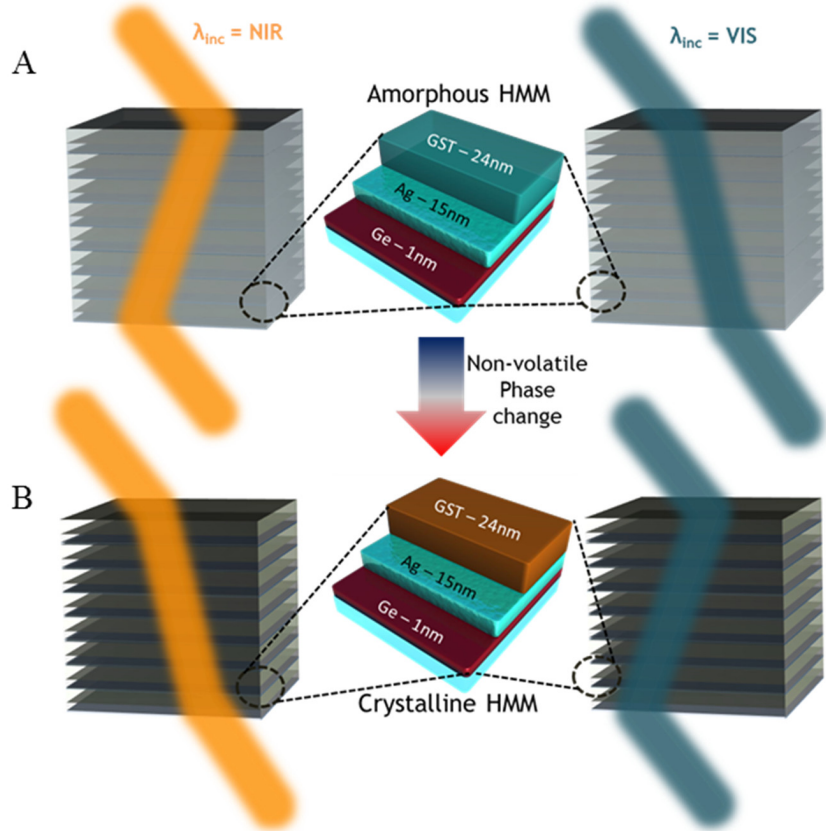

**Figure 1.** Schematic of the chalcogenide-based hyperbolic metamaterial in the (A) as-deposited configuration with the GST layers in the amorphous phase, and (B) after crystallization of GST layers by thermal annealing. In the former configuration, the HMM shows negative refraction for incident near-IR light and positive refraction for incident visible light, while in the latter, the situation is reversed by thermal annealing of the metamaterial. The HMM is comprised of seven periods of Ge, Ag and GST layers.

One of the most common approaches employed to realize an HMM is to fabricate a multilayer stack composed of alternate metal and dielectric layers of subwavelength thicknesses as shown schematically in Figure 1. In the effective medium limit, the dielectric permittivity tensor of such a layered nano-composite takes the form, $\varepsilon(r) = diag(\varepsilon_{xx}, \varepsilon_{yy}, \varepsilon_{zz})$, where $\varepsilon_{xx} = \varepsilon_{yy} = \varepsilon_\parallel$ and $\varepsilon_{zz} = \varepsilon_\perp$ are the effective dielectric constants of the structure in the direction parallel and perpendicular to the plane of the layers, respectively. Depending on the sign of the effective dielectric constants, the dispersion relation for light interacting with the HMM can be elliptical ($\varepsilon_\perp, \varepsilon_\parallel > 0$), hyperbolic Type – I ($\varepsilon_\perp, \varepsilon_\parallel > 0$) or hyperbolic Type – II ($\varepsilon_\perp > 0, \varepsilon_\parallel < 0$).

In our experiments, the HMM comprises of 7 periods of alternating layers of silver (Ag) and GST of thicknesses 15 nm and 24 nm, respectively, as shown schematically in Figure 1. An ultrathin layer (~ 1 nm thick) of germanium (Ge) is used as a wetting layer in order to ensure good quality silver films. The HMM was deposited on optically flat quartz substrates by RF sputtering (Kurt J. Lesker Nano 38). A base pressure of 5 x $10^{-5}$ mbar is achieved prior to deposition and high-purity argon is used as the sputtering gas. The substrate is held within 10 K of room temperature on a rotating platen 150 mm from the target to produce low-stress as-deposited amorphous films. The layer thicknesses were ascertained by means of a quartz crystal thickness monitor whose parameters for Ag, GST and Ge were calibrated based on stylus profilometer measurements on the respective single layers deposited separately.

Optical characterization of the HMM was performed by carrying out spectroscopic ellipsometric measurements using a J. A. Woollam M2000 ellipsometer, over a broad spectral range of 250 – 1500 nm, at multiple angles. The experimental data were analyzed using a uniaxial anisotropic model to extract the effective in-plane ($\varepsilon_\parallel$) and out-of-plane ($\varepsilon_\perp$) dielectric constants [4,28] of the HMM. The advantage of employing such a model is that it treats the HMM as an optically anisotropic 'blackbox' and enables understanding of the HMM's optical response without the knowledge of the dielectric constants of the individual layers composing the HMM.

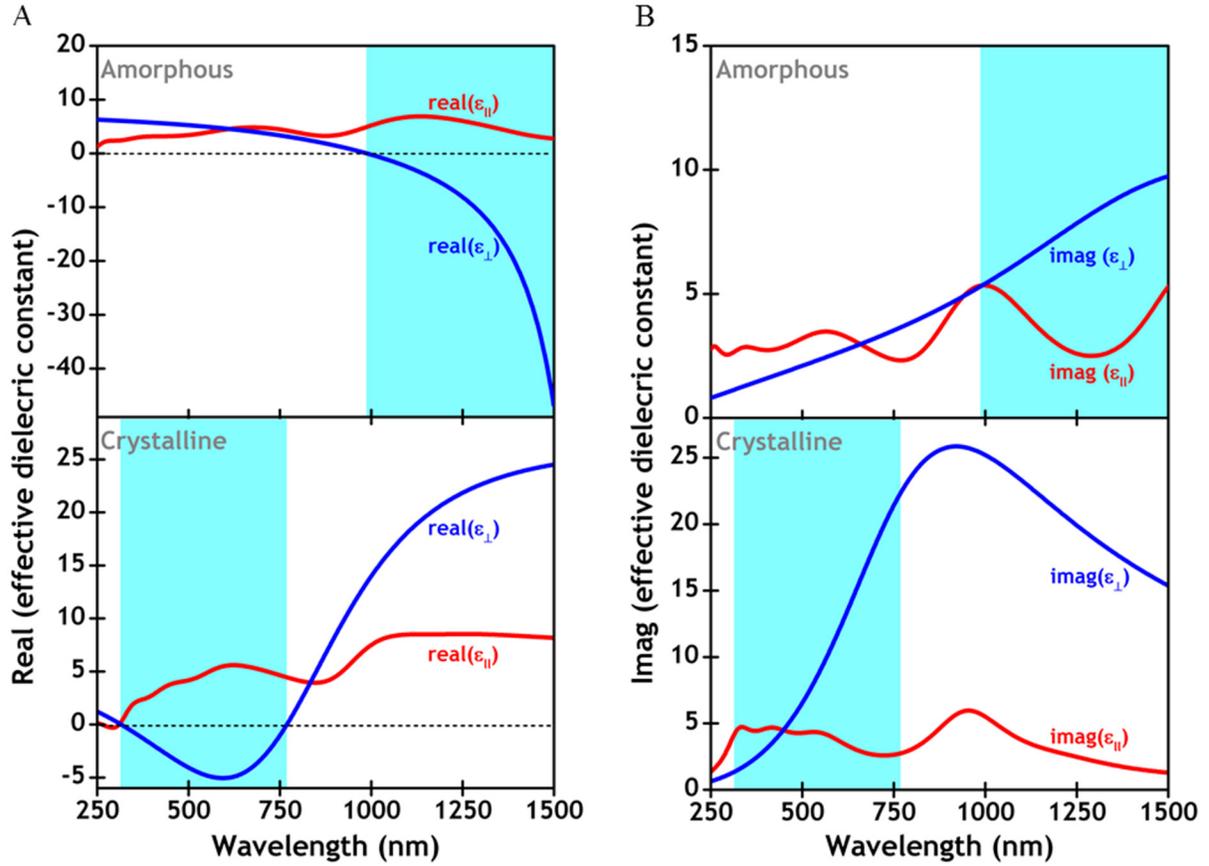

**Figure 2.** (A) Real and (B) imaginary parts of the effective dielectric constants $\varepsilon_\parallel$ and $\varepsilon_\perp$ of the HMM extracted from ellipsometry, with GST layers in the amorphous (top panel) and crystalline (bottom panel) phases. The shaded part indicates the region of hyperbolic dispersion.

The top panels of Figures 2A and 2B show the real and imaginary parts of effective dielectric constants of our HMM in the as-deposited configuration, with the GST layers in the amorphous phase. The structure exhibits Type-I hyperbolic dispersion ($\varepsilon_\perp < 0$, $\varepsilon_\parallel > 0$) in the near-IR region and elliptical dispersion ($\varepsilon_\perp, \varepsilon_\parallel > 0$) in the visible region. Subsequently, the HMM is annealed above its glass-transition point $T_g$ but below its melting point $T_m$ (around 110 and 630°C respectively) [29], to convert the GST phase from amorphous to crystalline. Real and imaginary parts of the effective dielectric constants of the annealed nanocomposite extracted from ellipsometry measurements are shown in the bottom panels of Figures 2A and 2B, respectively; with GST layers in the crystalline phase, the regime of hyperbolic dispersion gets switched – the HMM exhibits Type-I hyperbolic dispersion in the visible region and elliptical dispersion in the near-IR region.

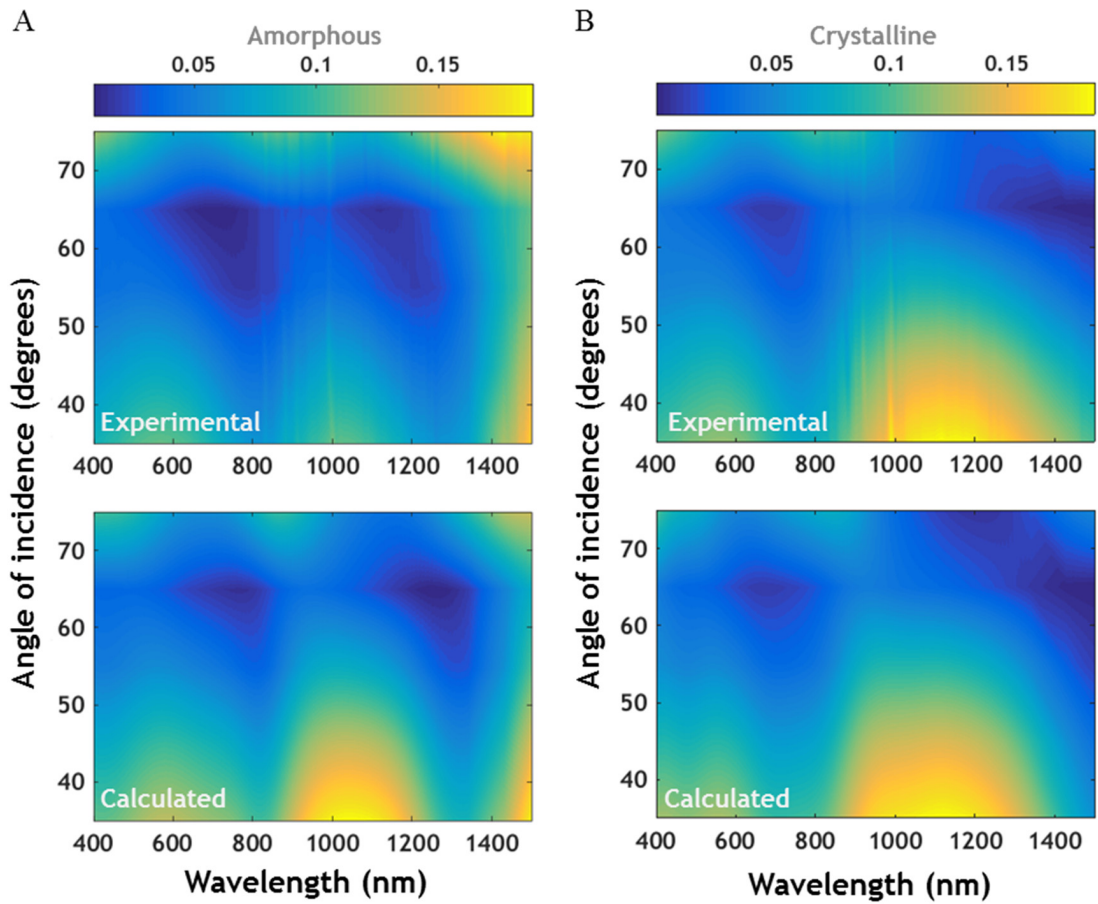

**Figure 3.** Experimental (top panel) and calculated (bottom panel) reflection spectra as a function of angle for TM polarized light incident on the HMM, with GST layers in the (A) amorphous, and (B) crystalline phases.

The HMMs in both phases were characterized further by carrying out polarized reflection measurements as a function of incidence angle. We compared the results of these measurements with the reflectivity spectra calculated based on a uniaxially anisotropic model, using the effective dielectric constants in figure 2. Figures 3A and 3B show the good agreement between the measured and calculated reflection spectra for TM polarization, for both as-deposited and annealed HMM samples. Similar measurements with TE polarized light also showed good agreement between the experimental and calculated spectra (see Supplementary Information). This confirms that the techniques as well as the anisotropic ellipsometry model we use to retrieve the effective dielectric constants of the HMMs in both phases are indeed accurate and describes their optical response quite well.

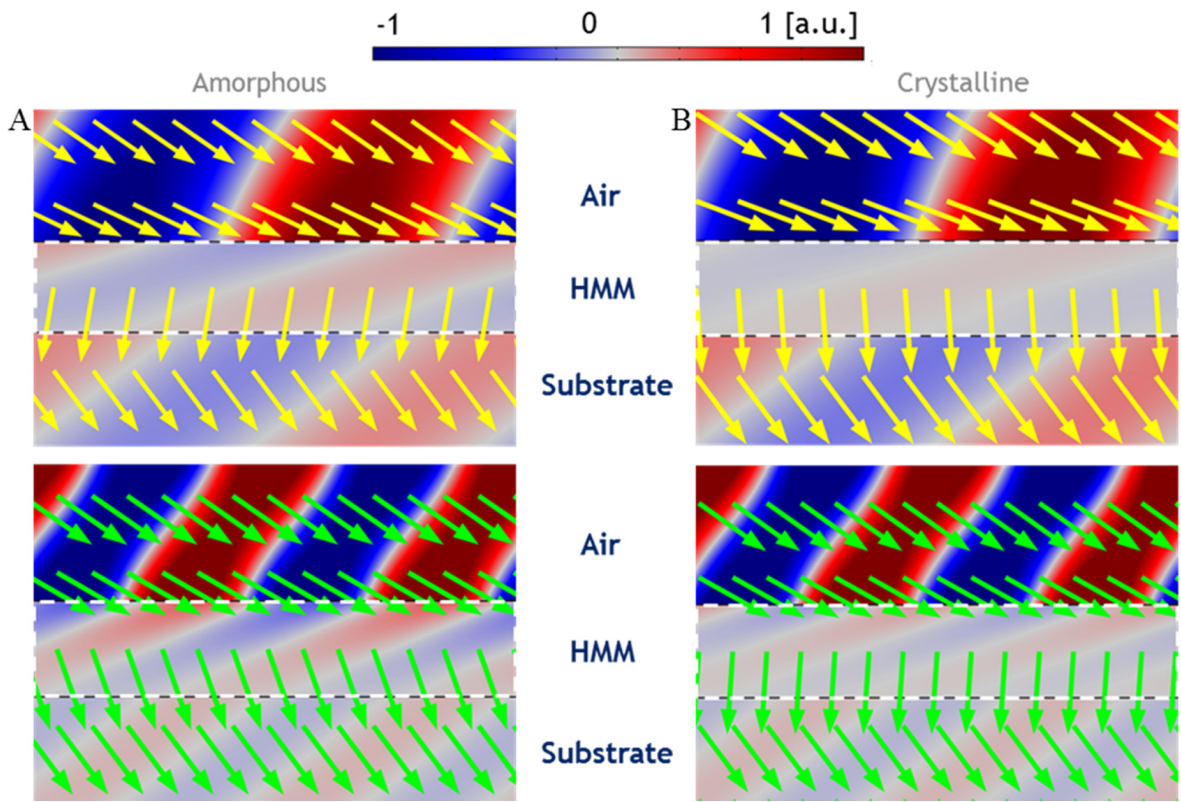

**Figure 4.** Color map of the electric field amplitude and the direction of power flow for TM polarized plane wave incident at an angle of 60° on the (A) Amorphous HMM and (B) Crystalline HMM, at wavelengths 1200 nm (top panel, yellow arrows) and 600 nm (bottom panel, green arrows). The HMM is highlighted by white dashed lines. Spectral regime of negative refraction is switched from NIR to visible by switching the phase of the GST layers from amorphous to crystalline. The simulations were performed using the experimentally determined dielectric constants of the HMM.

An intriguing aspect of materials showing Type – I hyperbolic dispersion is that they also show negative refraction of the Poynting vector [3,4]. This stems from the fact that the component of the Poynting vector in the direction parallel to the layers, of the beam refracted at the interface of the HMM, is decided by the sign of $\varepsilon_\perp$ [1,2]. We carried out finite element method (FEM) analysis of the HMM structure and simulated the electric field map and the Poynting vector for the HMM in the amorphous and crystalline phases. In the simulation, a plane wave with TM polarization is incident on the air-HMM interface at an angle of 60°. If the wavelength of the incident wave is 1200 nm (top panel), it experiences hyperbolic dispersion with $\varepsilon_\perp < 0$, which leads to sign reversal of the Poynting vector component parallel to the interface and therefore is refracted negatively, as shown in figure 4A. This is evident from the direction of the Poynting vector as well as the phase reversal of the beam refracted into the HMM. On the other hand, if the incident light has a wavelength of 600 nm (bottom panel), it does not experience hyperbolic dispersion and is positively refracted. This situation is inverted after conversion of the GST layers to the crystalline phase, where the hyperbolic dispersion regime is switched to the visible. The HMM now exhibits negative refraction at 600 nm and positive refraction at 1200 nm (figure 4B).

In conclusion, the work presented here is the first proof-of-concept demonstration of a chalcogenide-based multilayered metal-dielectric nanocomposite system and provides a platform to develop hyperbolic metamaterials capable of non-volatile switching, enabling access to hyperbolic and elliptical dispersion regions. As a proof of concept of the switchable nature of the hyperbolic response, we show that the HMM supports the co-existence of both positive and negative refraction in the visible and near-IR spectral regions, respectively; the sign of refraction in both spectral regions can be inverted by simply switching the phase of the chalcogenide glass.

**Acknowledgments**

This research was supported by the Singapore Ministry of Education (MOE2011-T3-1-005). The authors are grateful to Dr. Giorgio Adamo for useful discussions and assistance with structural characterization.

# Supplementary Information

## S1. Response of the HMM to TE polarized light

As in the case of TM polarized light, the response of the HMM to TE polarized light was also measured experimentally as a function of angle. Besides, using the dielectric constants shown in figure 2, we calculated the angle-resolved reflected spectrum for comparison. Figure S1 shows the good agreement between the experimental and calculated reflection spectra.

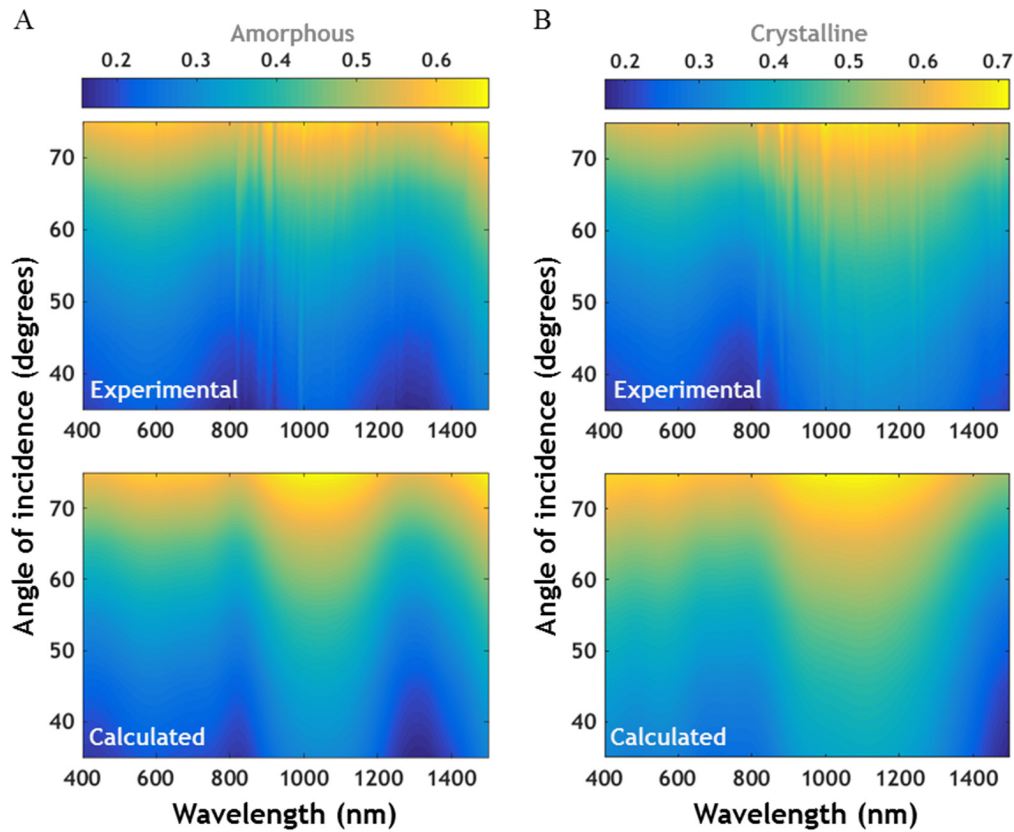

**Figure S1.** Experimental (top panel) and calculated (bottom panel) reflection spectra as a function of angle for TE polarized light incident on the HMM, with GST layers in the (A) amorphous, and (B) crystalline phases.